\documentstyle[12pt,titlepage]{article} 
\newcommand{\ba}{\begin{array}}
\newcommand{\ea}{\end{array}}
\newcommand{\bd}{\begin{displaymath}}
\newcommand{\ed}{\end{displaymath}}
\newcommand{\be}{\begin{equation}}
\newcommand{\ee}{\end{equation}}
\newcommand{\bea}{\begin{eqnarray}}
\newcommand{\eea}{\end{eqnarray}}

\begin{document}
\begin{titlepage}
\vspace*{0.5truein}

\begin{flushright}
\begin{tabular}{l}
MRI-PHY/5/96\\
January, 1996
\end{tabular}
\end{flushright}
\vskip .6cm

\begin{center}
{\Large\bf DISTINCTIVE SIGNALS OF SPONTANEOUS R-PARITY BREAKING  
AT LEP-II}\\
{\large Rathin Adhikari$^1$\\}
Theoretical Physics Division    \\
Physical Research Laboratory,
Navrangpura, Ahmedabad - 380 009, INDIA \\
{\large Biswarup Mukhopadhyaya$^2$\\}
Mehta Research Institute,
10 Kasturba Gandhi Marg,
Allahabad - 211 002, INDIA \\
\end{center}
\vskip .5cm

\begin{center}
{\bf ABSTRACT}\\
\end{center}

\noindent We consider the signals of pair-produced charginos in an
$e^{+}e^{-}$ collider, in a scenario where R-parity is spontaneously
broken. The possibility of lepton-chargino mixing, together with the
presence of a Majoron, opens up the two-body decay channel of a chargino
into a tau and a Majoron. We have studied the regions where this is the
dominant decay of a chargino, and have shown that over a large region of the
parameter space the ensuing signals, namely tau-pairs with missing-$p_T$, 
may rise well
above standard model backgrounds at LEP-II. Such signals can also enable 
one to distinguish between spontaneous and explicit R-parity breaking.

\vskip .5in

\noindent
$^{1}$E-mail : ~~rathin@prl.ernet.in  \\ 
$^{2}$E-mail :biswarup@mri.ernet.in

\end{titlepage}

\textheight=8.9in

Supersymmetric (SUSY) theories which do not conserve R-parity, defined as 
$R =  (-1)^{3B+L+2S}$, have already received considerable attention \cite{1}. 
All particles in the standard model have $R = 1$ whereas their superpartners
have $R = -1$. It also follows from the above definition that the
nonconservation of either baryon or lepton number can lead to R-parity
violation. While the former is rather seriously restricted by, for example,
the absence of proton decay, breakdown of the latter is still
phenomenologically viable. Thus an R-parity breaking theory may often be
looked upon as a SUSY theory where lepton number is not conserved. Such a
scenario implies a wide variety of new phenomena that can be looked for in
the up-and coming experiments.

R-parity can be broken in two ways-- explicitly or spontaneously. The former
option consists in introducing L(or B?)-violating interactions by hand in the
superpotential. Such a model is blessed with the same economy as that of the
minimal supersymmetric standard model (MSSM) and its predictions are 
well-studied \cite{2}.  The latter possibility entails the breakdown of R-parity
through the vacuum expectation value (VEV) acquired by some scalar fields
possessing non-zero lepton numbers. Though such a scenario results in an
augmentation of the particle spectrum, it has some theoretical advantages
as well. For example, in a situation where the VEV of an SU(2) singlet
scalar breaks R-parity, there can be a natural origin of light sterile
neutrinos \cite{3} whose existence is being widely conjectured to explain 
various experimental results.

In this paper, we propose to concentrate on some observable signals of 
spontaneously broken R-parity. An immediate result of spontaneous breakdown 
of lepton number in this case is the appearance of a massless Goldstone boson
which is called the Majoron \cite{4}. Such a Majoron can be produced in the 
decay of a chargino, together with a charged lepton like the tau. We give here
a detailed prediction of the signals of such decays at the LEP-II experiments.

Two types of models have been proposed in recent approaches to spontaneous
R-parity breaking. Many of their phenomenological implications have also been 
studied. In the first of these \cite{5}-\cite{7}, the additional ingredients 
over and above
those in the MSSM are the two isosinglet neutral superfields $\nu^c_i (i=1-3)$
and $S$ carrying lepton numbers -1 and 2 respectively. The leptonic part of
the superpotential here is of the form

\begin{equation}
W = {\mu} H_{1} H_{2} + {\lambda_l} {l^c} L H_1 + {\lambda_\nu} {\nu^c} L H_2
+ {\lambda_S} S \nu^c \nu^c
\end{equation}

\noindent where $L$ stands for the left chiral lepton-neutrino doublet 
and $l$, the
right chiral charged leptons.  $H_1$ and $H_2$ are the Higgs doublets
giving masses to the down-and up-type quarks respectively. Here the VEV's
of the scalars $\tilde{\nu^c}$ and $\tilde{S}$ (as well as that of 
$\tilde{\nu_L}$) can break $L$ spontaneously. 

In the other model \cite{8}-\cite{13}, three kinds of neutral singlet 
superfields ($\phi$, $\nu^c$, S) are incorporated with lepton numbers 
(0, -1, 1). The corresponding superpotential is, in the same notation as above,

\begin{equation}
W = h_0 \phi H_{1} H_{2} + {\lambda_l} {l^c} L H_1 + 
{\lambda_\nu} {\nu^c} L H_2 + {\lambda_S} S \nu^c \phi
\end{equation}

Again, lepton number is violated by $<\tilde{\nu}_{R,L}>$ and $\tilde{S}$.
This second possibility offers a natural scale to the Higgsino mass
parameter $\mu$ through the VEV of $\phi$, although it requires a larger 
particle spectrum.

Here we shall focus on features which are common to both of the above models,
namely:

\noindent (i) After spontaneous symmetry breaking, one is left with terms of
the form $LH_2$. This gives rise to mixing between charginos and charged 
leptons  (as also between neutralinos and neutrinos). Note that
although such terms may apparently be rotated away by redefinition of the
$L$  and $H_2$ fields, the mixing inevitably shows up in the scalar potential
\cite{14}.

\noindent (ii) A massless Goldstone boson, named the Majoron $(J)$, emerges
on the scene. The field $J$ is composed dominantly of singlet scalars, but 
carries a doublet component also (because $<\tilde{\nu_{L}}>$
in general contributes to symmetry breaking). However, constraints from
$Z-$width measurements \cite{7},\cite{9} as well as those from the cooling of 
red giant stars \cite{15}
require its $\tilde{\nu_{L}}$-component to be quite small.

When chargino-lepton mixing takes place, there is a (5 X 5) mass matrix to
be diagonalised in that sector. The above features cause the Majoron to have
tree-level coupling with a chargino (i.e. one of the two heaviest physical
states) and a charged lepton (i.e. one of the three lightest physical states).
This can be understood if we remember that in the weak eigenstate basis,
the lepton fields do not couple to the Majoron which is dominantly an SU(2)
singlet. However, the charginos do couple to it through their Higgsino
components. Such a disparity of couplings  prevents the 
Glashow-Illiopoulos-Maiani (GIM) mechanism from being operative when 
the mass matrix is diagonalised, thereby connecting different mass eigenstates
in the Majoron interaction.

The most general parametrisation of such an interaction is

\begin{eqnarray}
{\cal L}_{Jl\chi} = {{J}\over{\sqrt{2}}} {\chi_i} 
\left[ A_1{ {1 - \gamma_5}\over{2}} + A_2{ {1 + \gamma_5}\over{2}}\right] l_j
\end{eqnarray}

There are strong experimental constraints on lepton 
flavour violating processes involving the electron or the muon. 
Also, difficulties related to baryogenesis can be avoided if at least one
family has no lepton flavour violation \cite{16}. This makes it natural to 
assume that R-parity violating effects of the type discussed above are 
suppressed  excepting for the tau family. In the remaining discussion we 
shall assume that only the tau mixes with the charginos.

Neglecting the $\tau$-mass, the width for 
$\chi^{\pm} \longrightarrow \tau^{\pm} J$ is given by

\begin{eqnarray}
\Gamma(\chi \longrightarrow \tau J) = {{{m_\chi}A^2}\over{32 \pi}}
\end{eqnarray}

\noindent  Where $A^2 = A^2_1 + A^2_2$. The
parameter $A$ will henceforth be used as a measure of the lepton number violating
interaction strength. The laboratory constraints on the $\tau$-neutrino mass
{\it viz.} $m_{\nu_\tau} < 31$ MeV is perhaps the strongest source of 
constraint on the parameter $A$. However, the actual limit
depends on the actual 
superpotential as well as on quantities such as the parameter $\mu$ and the 
right-handed sneutrino VEV. Approximate estimates based on the simplified 
assumption that all mass parameters are about 100 GeV implies \cite{7} that  
$A$ can be as high as  a few times $10^{-2}$. We shall treat $A$ as a free 
parameter here.

If the charged sleptons and sneutrinos are heavier than the lighter 
chargino, then the only decay channels available to the latter are
(i) the R-parity violating decay
$\chi^{\pm} \longrightarrow \tau^{\pm} J$, and (ii) the three-body decay
$\chi^{\pm} \longrightarrow \chi_0 f \bar{f'}$ ($\chi_0$ being the lightest
neutralino), which is allowed in MSSM itself. In addition, there may be
R-parity violating three-body decay modes of the type 
$\chi^{\pm} \longrightarrow \tau^{\pm} \nu \bar{\nu}$. These will be suppressed
by the R-violating parameter, and can be neglected when one is concerned
with the branching ratio for the two-body decay. The relative strengths of
(i) and (ii) depend on the quantity $A$ in equation 4, and on the choice
of the other SUSY parameters (gluino mass, $\mu$, tan$\beta$ = $<{H_2}/{H_1}>$)
which  fix the masses and couplings governing the R-parity conserving decays.
It may be remarked  that as far as  R-conserving processes are concerned,
the relevant interactions and mass relationships have negligibly small
effects from R-violating terms. Hence the widths for such decays 
are equal to those obtained in the MSSM to a high degree of accuracy.  

Since the Majoron is massless and sterile, 
$\chi^{\pm} \longrightarrow \tau^{\pm} J$ leads to signals of the form
$\tau^{+} \tau^{-} + \not{p_T}$ in $e^{+} e^{-}$ colliders such as the
LEP-II where the lighter chargino can be pair-produced. Such signals arise
in the standard model from $W-$pairs. However, the decay 
$W {\longrightarrow} \tau \bar{\nu_\tau}$ has a branching ratio of about 11
per cent, and thus tau-pair production there suffers from a net suppresion
of two orders of magnitude. There is no such suppression in the region of the
parameter space where $\chi^{\pm} \longrightarrow \tau^{\pm} J$ is dominant.
Therefore, in such a region the
$\tau$-pair production rate can considerably exceed what is predicted from
$W-$decays. Given the fact that tau-identification efficiency can easily be as 
high as 80 per cent at LEP-II, those tau-pairs may prove to be the most 
important signatures of spontaneous R-breaking if charginos are light enough to
be pair-produced.

Chargino pair-production in $e^{+} e^{-}$ annihilation proceeds through
s-channel diagrams mediated by the photon and the $Z$, and
through a $t-$channel 
diagram mediated by the sneutrino \cite{17}. The latter is dominant at 
LEP-II energies
so long as $m_{\tilde \nu} \leq$100-120 GeV. To calculate the branching ratio
for $\chi^{\pm} \longrightarrow \tau^{\pm} J$, one needs to obtain also the 
three-body decay widths. That involves the computation of diagrams with the
$W$, sleptons and squarks in the propagators. We assume all squarks and all 
sleptons (charged and neutral) to be separately degenerate.

The cross-sections for 
$e^{+} e^{-} \longrightarrow \chi \chi \longrightarrow \tau \tau JJ$ have
been calculated here at LEP-II energies. For various choices of the 
SUSY parameters, we have made a full calculation of the different deecay
widths of the chargino and accordingly fixed the branching ratio for the 
R-violating two-body decay. The event selection criteria here are 
$E_\tau \ge 5$ GeV and $\not{p_T} \ge$ 5 GeV. Since the 
energy and ${p_T}$ available from
$\chi^{\pm}$-decay are rougly equipartitioned between the tau and the
Majoron, it is found that the signals under question are practically 
unaffected by these cuts. For this reason, the spin-averaged 
cross-sections for the said final states modulo the suggested cuts agree with 
the spin-correlated ones. 

In figures 1-4, we show the $\tau \tau JJ$ cross-sections plotted against
the parameter $A$ for different values
of the mass of the lighter chargino, using two values each of the gluino mass
($m_{\tilde{g}}$) and tan$\beta$. These are sample results; for higher 
tan$\beta$, the results are in general more optimistic. The squark mass has 
been fixed at 200 GeV.
A slepton mass of 100 GeV is used in figures 2-4; only in figure 1 it
is 80 GeV. A comparison between figures 1 and 2 shows the sensitivity of
our predictions to the slepton mass. The enhancement due to a lower slepton
mass can be attributed to the $t-$channel propagator effect
in the production cross-section. Also, the cross-section gradually
flattens out as the parameter $A$ increses beyond about 0.01. This is because
then onwards, the two-body decay of the chargino is completely dominant,
and the branching ratio remains almost 100\% throuhgout.  

    The main standard model background {\it i.e.} 
$e^{+} e^{-} \longrightarrow W W \longrightarrow \tau \tau + \not{p_T}$ 
has a cross-section of about 0.18 pb at LEP-II. The other background, namely
that from a $Z-$pair, is even more suppressed by the branching
ratios for leptonic 
$Z-$decay. Our results clearly show
that over a rather sizable range of the parameter space the chargino
decay signals are larger than such backgounds by upto one order of magnitude
in the parameter range shown here. Although 
the two kinds of event topologies are very similar, it is easy to discern
the Majoron signals from a shower of tau-pairs with a large missing
$p_T$. The tau's are separated into two opposite hemispheres so that the
narrow, low-multiplicity jets arising from them can be easily distinguished.

 Now consider the strengths of similar signals in the MSSM and in a scenario
 with  explicit R-parity breaking. In the MSSM case, a chargino  could
give rise to a tau together with the lightest supersymmetric
particle (LSP) and a tau-antineutrino. The chargino pair-production rates would
be the same as in our case here. However, the decay into a tau 
in the MSSM is controlled by the gauge coupling which treats all the lepton
flavours similarly. Thus, for a pair of tau's to be produced, a minimum 
branching fraction of 1/3 on each side is unavoidable. Also, there
are the ($q \bar{q'} + LSP$) final states which eat further into the available
decay rates unless the squarks are very massive. This means that the MSSM
signals will be small compared to the case 
under study here. Similarly, with explicit
breakdown of R-parity, $\tau^{\pm} + \not{p_T}$ can occur from three-body
decays of chargino and neutralino pairs. Again, all the three lepton flavours
are simultaneously produced, in addition to the quark final states, thereby
reducing the branching fraction for tau-pairs. For spontaneous R-breaking,
on the other hand, the $\tau$-coupling of a chargino and a Majoron is 
favoured,  from considerations that we have already stated. The tau-signal in
the latter case, 
therefore, may enable one to distinguish spontaneous R-parity breaking from
explicit R-violation.

    Similar types of events would also result from pair-produced
charged Higgs bosons which can have the
$\tau - \nu_{\tau}$ decay mode. However, with a centre-of-mass energy of 
180 GeV, the pair production cross-section is approximately 0.1 pb for a 
charged Higgs pair of mass 80 GeV \cite{18}. Even for a theoretically 
unfavourable charged Higgs
of mass 70 GeV it is not more than 0.25 pb. Given the fact that there is a
further overall branching fraction suprression of 0.25 for going into a
tau on each side, this final state cannot rise above the $WW-$backgorund, 
and is expected to be drowned by the Majoron channel when the latter is
available.

    In conclusion, we have scanned the parameter space to demonstrate 
 that the production of tau-pairs
 plus missing transverse momentum at LEP-II can be considerably enhanced
 in a scenario with spontaneously broken R-parity. This should serve
 as a distinct experimental signal for this kind of models, as compared to 
 both the MSSM and an explicitly broken R-parity, during the
 operation of LEP-II.
 
 Acknowledgements: We thank A. Joshipura and A. Datta for helpful discussions.
 R.A. thanks the Mehta Research Instituite for hospitality during the period
 when this work was done.

\newpage
\centerline {\large {\bf Figure Captions}}

\hspace*{\fill}

\hspace*{\fill}

\noindent Figure 1: 

\noindent
$\sigma(e^{+}e^{-} \longrightarrow \tau \tau J J$) against the parameter A,
with gluino mass = 250 GeV, tan$\beta$ = 10, squark mass = 200 GeV,
sneutrino mass = 80 GeV. The solid, dotted and dashed
lines correspond respectively to chargino mass = 60, 70 and 80 GeV.
$\sqrt{s}$ = 180 GeV. 

\vskip .25in

\noindent Figure 2: 

\noindent
Same as in figure 1, but with sneutrino mass = 100 GeV.

\vskip .25in

\noindent Figure 3: 

\noindent
Same as in figure 2, with gluino mass = 300 GeV. The solid, dotted and dashed
lines correspond respectively to chargino mass = 70, 80 and 90 GeV. 
$\sqrt{s}$ = 200 GeV.

\noindent Figure 4: 

\noindent
Same as in figure 2, with gluino mass = 250 GeV, tan$\beta$ = 2, chargino mass
= 90 GeV (solid line), and gluino mass = 300 GeV, tan$\beta$ = 2, chargino
mass = 70 GeV (dashed line). 
$\sqrt{s}$ = 190 GeV.

\end{document}